\documentclass[12pt]{article}
\usepackage{graphicx}
\usepackage{xcolor}
\usepackage{xspace}
\usepackage{amsmath}
\usepackage{dcolumn}

\newcommand\pubdate{\today}

\def\institute{Institut f\"ur Experimentalphysik, Universit\"at Hamburg\\
Luruper Chaussee 149, 22761 Hamburg, Germany}
\def\support{\footnote{Supported by the German Federal Ministry of Education and Research.}}

\def\Title#1{\begin{center} {\Large #1 } \end{center}}
\def\Author#1{\begin{center}{ \sc #1} \end{center}}
\def\Address#1{\begin{center}{ \it #1} \end{center}}

\newcommand\pubblock{\rightline{\pubdate}}
\newenvironment{Abstract}{\begin{quotation}  }{\end{quotation}}
\newenvironment{Presented}{\begin{quotation} \begin{center} 
             PRESENTED AT\end{center}\bigskip 
      \begin{center}\begin{large}}{\end{large}\end{center} \end{quotation}}

\usepackage{pennames-pazo}
\usepackage{ptdr-definitions}
\newcommand{\cmsSymbolFace}{\mathrm}

\newcolumntype{d}[1]{D{.}{.}{#1} }

\newcommand{\chisq}{{\ensuremath{\chi^2}}}

\newcolumntype{L}{>{$}l<{$}}
\newcolumntype{R}{>{$}r<{$}}
\newcolumntype{C}{>{$}c<{$}}

\newcolumntype{T}[2]{%
    >{\adjustbox{angle=#1,lap=\width-(#2)}\bgroup}%
    l%
    <{\egroup}%
}

\newcommand{\lind}[1]{_\mathrm{#1}}

\newcommand{\sub}[1]{{\ensuremath{\lind{#1}}}}
\newcommand{\subT}{{\sub{T}}}

\newcommand{\pT}  {\ensuremath{p\subT}\xspace}

\newcommand{\mtop}{\ensuremath{m_{\cPqt}}\xspace}
\newcommand{\mtfit}{\ensuremath{m_{\cPqt}^\text{fit}}\xspace}

\newcommand{\mWreco}{\ensuremath{m_{\PW}^\text{reco}}\xspace}
\newcommand{\Pgof}{\ensuremath{P_\text{gof}}\xspace}
\newcommand{\DRbb}{\ensuremath{\Delta R(\cPqb\cPaqb)}\xspace}
\newcommand{\JSF}{\ensuremath{\text{JSF}}\xspace}

\newcommand{\statJSF}{\ensuremath{\,\text{(stat+JSF)}}\xspace}

\newlength\cmsTabSkip

\begin{document}
\begin{titlepage}
\pubblock

\vfill
\Title{Top quark mass measurement in the \ttbar all-jets final state with the CMS experiment at $\sqrt{s}=13\TeV$}
\vfill
\Author{Johannes Lange\support{} on behalf of the CMS Collaboration}
\Address{\institute}
\vfill
\begin{Abstract}
The top quark mass is measured using $35.9~\mathrm{fb}^{-1}$ of LHC proton-proton collision data collected with the CMS detector at $\sqrt{s}=13~\mathrm{TeV}$ in 2016. The measurement uses the $\mathrm{t}\overline{\mathrm{t}}$ all-jets final state, which comprises a total of six jets. A kinematic fit is performed to reconstruct the decay of the $\mathrm{t}\overline{\mathrm{t}}$ system and suppress QCD multijet background. By means of the ideogram method, the top quark mass is determined, simultaneously constraining an additional jet energy scale factor ($\text{JSF}$). The result of $172.34\pm0.20\,\text{(stat+JSF)}\pm0.76\,\text{(syst)}~\mathrm{GeV}$ for the top quark mass is in good agreement with previous measurements in the same and different final states.
\end{Abstract}
\vfill
\begin{Presented}
$11^\mathrm{th}$ International Workshop on Top Quark Physics\\
Bad Neuenahr, Germany, September 16--21, 2018
\end{Presented}
\vfill
\end{titlepage}
\def\thefootnote{\fnsymbol{footnote}}
\setcounter{footnote}{0}

\section{Introduction}
A top quark mass (\mtop) measurement is presented using $35.9~\mathrm{fb}^{-1}$ of LHC proton-proton collision data collected with the CMS detector~\cite{Chatrchyan:2008aa} at $\sqrt{s}=13~\mathrm{TeV}$ in 2016.
Proton-proton collision events are selected in which a top quark-antiquark pair (\ttbar) is produced and
both daughter \PW~bosons decay to quark-antiquark pairs.
Thus, the final state consists of at least six jets.
A more complete description of the analysis can be found in Ref.~\cite{CMS:2018nbt}.
A similar analysis at $\sqrt{s}=13\TeV$ using the lepton+jets channel has been presented in Ref.~\cite{Sirunyan:2018gqx}.

\section{Event selection and simulation}

Jets are required to fulfill $\pT>30\GeV$ and $\abs{\eta}<2.4$.
Events are selected using a trigger that requires at least six jets with $\pT>40\GeV$, $\HT=\sum_\text{jets} \pT > 450\GeV$, and at least one b~jet.
In the offline selection, the same requirements are applied.
In addition, a second b~jet is required, and the separation of the two b~jets must fulfill $\DRbb>2.0$.

Signal \ttbar events are simulated using \POWHEG~v2~\cite{Nason:2004rx,Frixione:2007vw,Alioli:2010xd,Campbell:2014kua}
as matrix-element generator, interfaced with \PYTHIA~8.219~\cite{Sjostrand:2007gs} for the parton shower and hadronization.
The detector response is modeled using \GEANTfour~\cite{Agostinelli:2002hh}.

\section{Kinematic fit and background estimation}

To select the correct jet-to-parton assignment for each event and to improve the mass resolution, a kinematic fit is applied.
For each assignment, the jet momenta are varied to fulfill three constraints.
The invariant masses of the jet pairs belonging to the parent \PW~boson are required to be $80.4\GeV$ and the invariant masses of both top quark candidates are forced to be equal.
For each assignment
\begin{equation*}
  \chisq = \sum_{j\in \text{jets}} \left[
  \frac{\left(\pT_j^\text{reco} - \pT_j^\text{fit}\right)^2}{\sigma_{\pT_j}^2} +
  \frac{\left(\eta_j^\text{reco} - \eta_j^\text{fit}\right)^2}{\sigma_{\eta_j}^2} +
  \frac{\left(\phi_j^\text{reco} - \phi_j^\text{fit}\right)^2}{\sigma_{\phi_j}^2}\right]
\end{equation*}
is minimized and only the assignment giving the lowest \chisq~value is considered in the following.
Quantities labeled ``reco'' refer to the originally reconstructed jets, while ``fit'' labels the momenta
returned by the fit.
The \chisq~probability (\Pgof) is required to be greater than 0.1 for events to be accepted.

\begin{figure}[ht!]
  \centering
  \includegraphics[width=.5\linewidth]{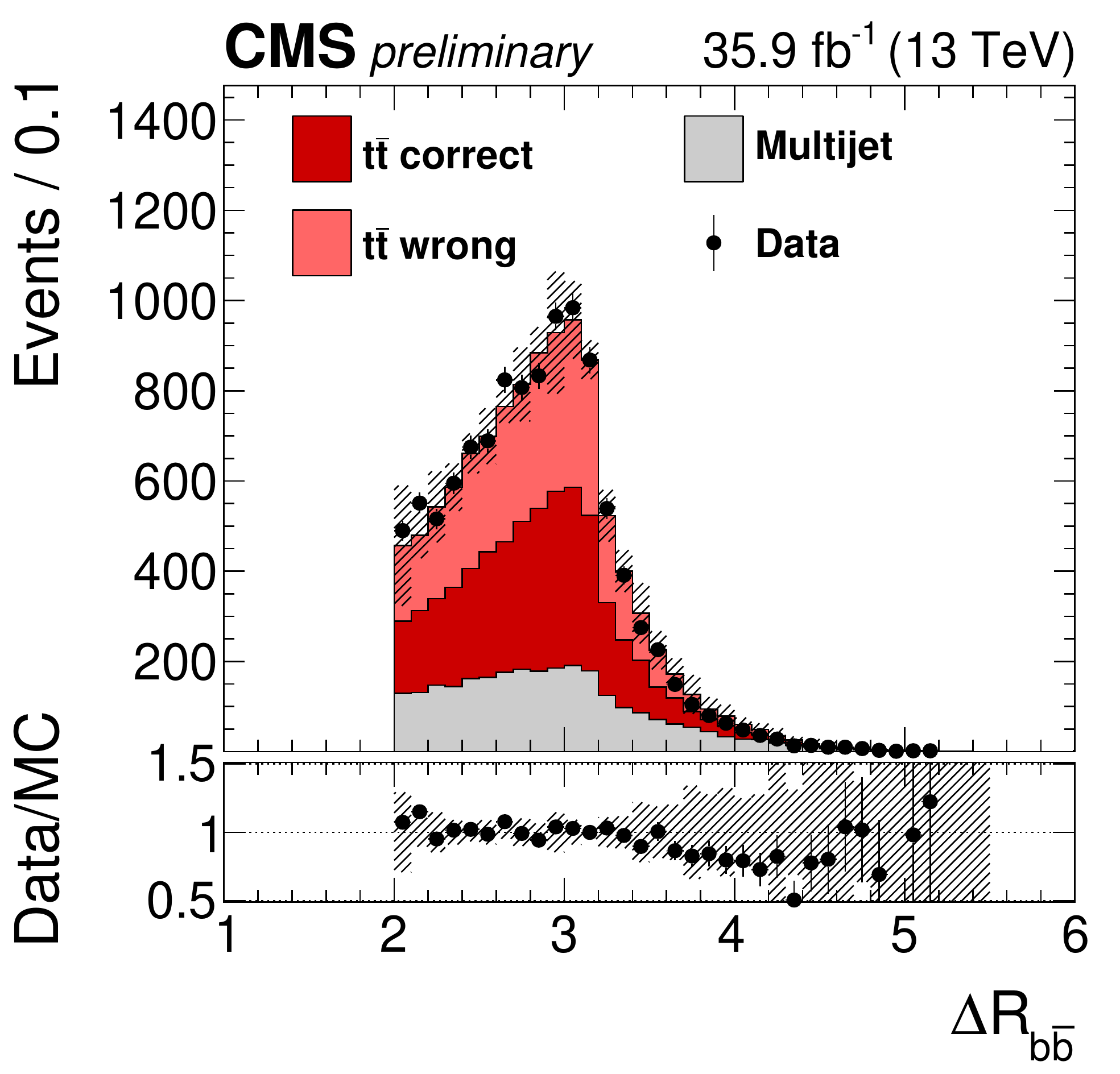}%
  \includegraphics[width=.5\linewidth]{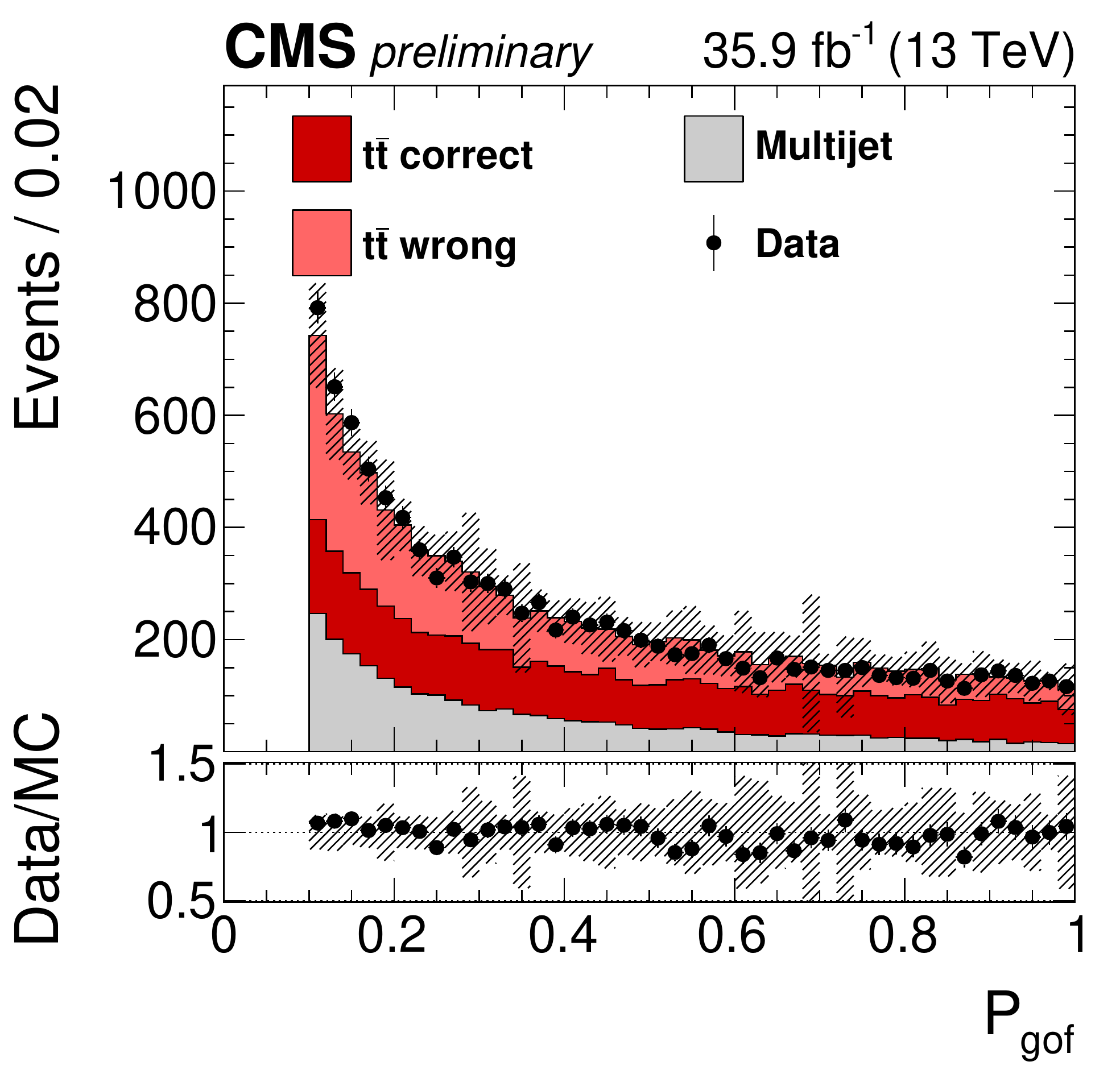}\\
  \includegraphics[width=.5\linewidth]{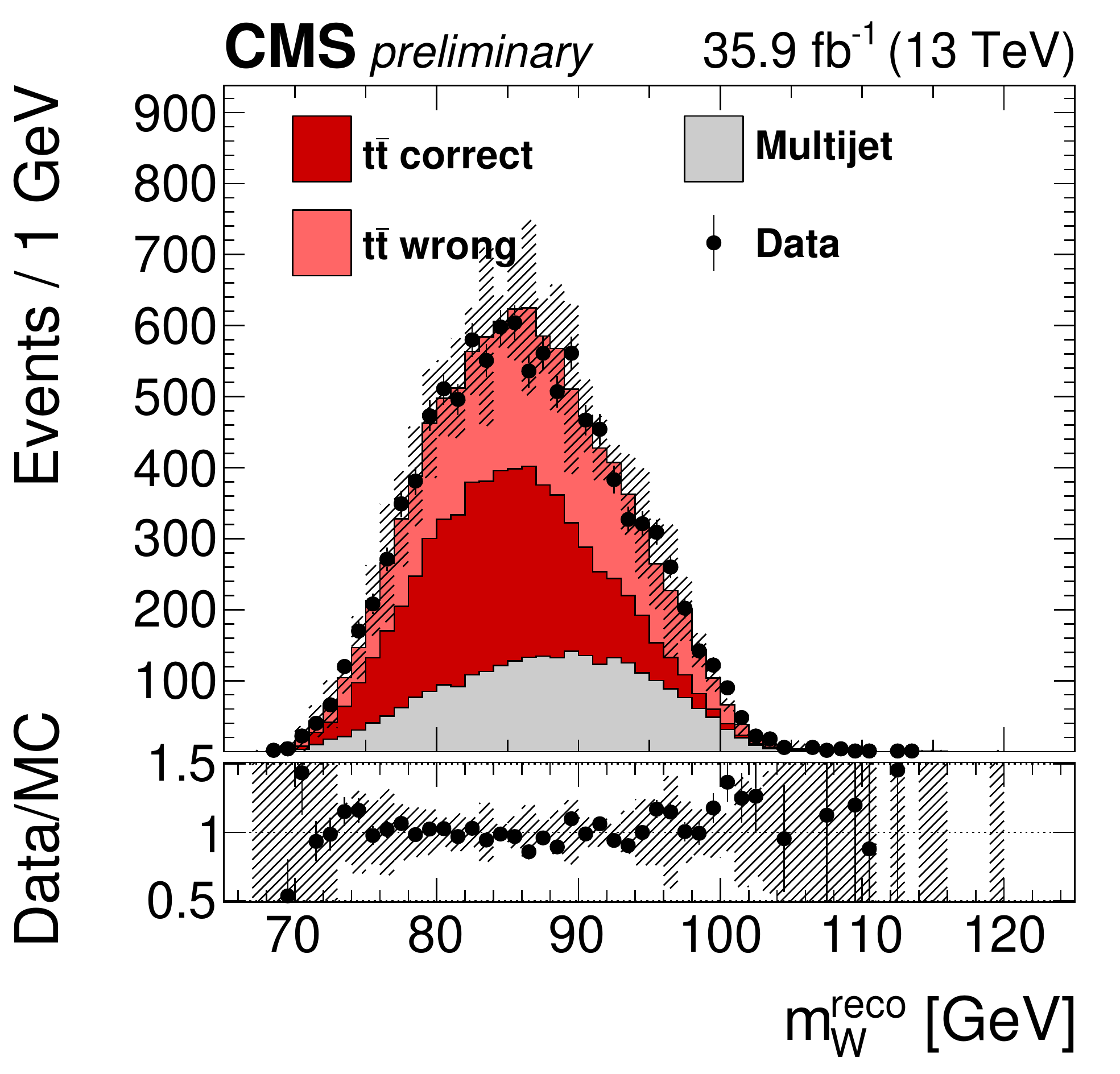}%
  \includegraphics[width=.5\linewidth]{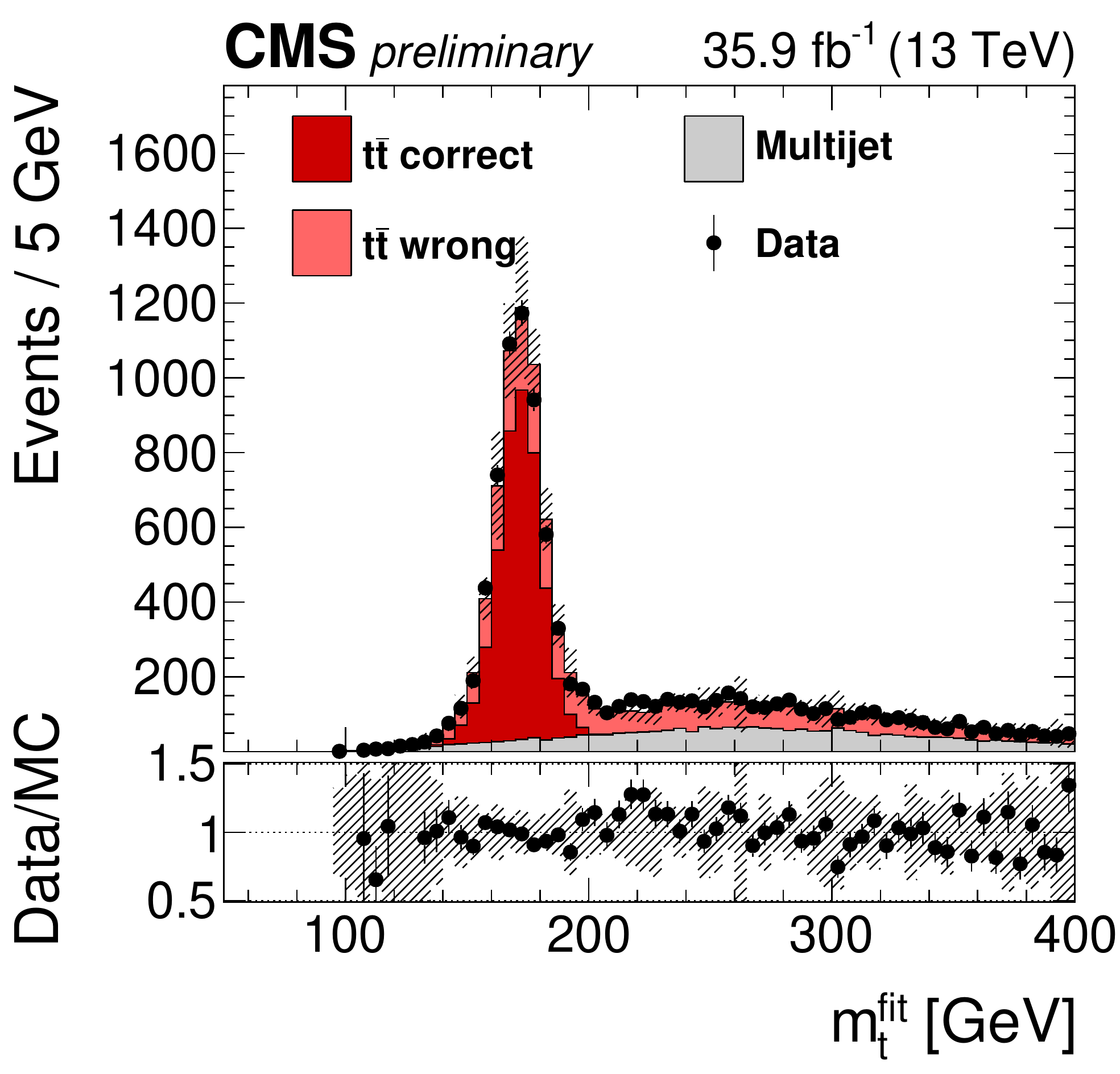}%
  \caption{The \DRbb (top left), \Pgof (top right),
    fitted top quark mass (bottom left), and reconstructed \PW~boson mass (bottom right) distributions of data
    compared to signal simulation and the multijet background estimate.
    The shown reconstructed \PW~boson mass is the average of both \PW~bosons in the event.
    The hashed bands represent the total uncertainty of the prediction~\cite{CMS:2018nbt}.
  \label{fig:distributions}}
\end{figure}

Background arises from QCD mutijet production. Owing to the large production cross section, a substantial amount of multijet events can pass the kinematic fit selection.
In these events, the topology imposed by the fit is met just by combinatorial chance, which does not depend on the b~jet multiplicity.
Therefore, the shape of the background distribution is estimated from data events containing exactly zero b~jets, and validated using simulated events.
The normalization is initially unknown and a free parameter in the \mtop extraction described below.
The \DRbb, \Pgof, fitted top quark mass, and reconstructed \PW~boson mass distributions are shown in Fig.~\ref{fig:distributions}.
Here, the background estimate is normalized to the difference of selected data events and expected signal events.
Good agreement of data and prediction is observed for all distributions.

\section{Mass extraction}

The ideogram method~\cite{Abdallah:2008ad,Chatrchyan:2012cz} is used to extract \mtop, as well as an additional global
jet energy scale factor (\JSF).
The likelihood
\begin{equation*}
\mathcal{L}\left(\text{sample}|\mtop,\text{JSF}\right)
= \prod_{\text{events}}P\left(\text{event}|\mtop,\text{JSF}\right)
= \prod_{\text{events}}P\left(\mtfit,\mWreco|\mtop,\text{JSF}\right)
\end{equation*}
is maximized, optionally including a prior probability for the \JSF, \ie, maximizing
\begin{equation*}
P(\JSF) \cdot \mathcal{L}\left(\text{sample}|\mtop,\text{JSF}\right)\,.
\end{equation*}
Since \mWreco is sensitive to changes in the jet energy scale, systematic uncertainties affecting the jet energies can be constrained by the inclusion of the \JSF in the fit.
The probability densities $P$ are described by analytical functions depending on \mtop and \JSF.
Three versions of the maximum likelihood fit are performed: In the 1D measurement, the \JSF is fixed to unity, while it is a
free parameter in the 2D measurement.
The third (hybrid) measurement imposes a Gaussian constraint around unity for the \JSF.
To remove residual biases introduced by the method, pseudo-experiments are performed for different values of \mtop and \JSF and
a calibration curve is extracted, which is used in the following.

\section{Systematic uncertainties and results}

To determine systematic uncertainties, pseudo-experiments are performed using different systematic variations as input and
the differences with respect to the default sample are quoted.
The full description of the systematic uncertainty sources can be found in Ref.~\cite{CMS:2018nbt}.

The hybrid fit, constructed to provide the lowest total uncertainty, yields
\begin{align*}
  \mtop & = 172.34\pm0.20\statJSF\pm0.76\syst\GeV\,.
\end{align*}
All other results and a complete list of the different systematic uncertainties can be found in Ref.~\cite{CMS:2018nbt}.
The main systematic uncertainty of $0.36\GeV$ stems from the modeling of color reconnection, which has been estimated using
new models that have become available in \PYTHIA~8~\cite{Argyropoulos:2014zoa,Christiansen:2015yqa}.
The uncertainty determined using the new models is increased with respect the values quoted in CMS publications at lower center-of-mass energies.
The other uncertainties are of similar size as in previous measurements reported by the CMS Collaboration.

\section{Summary}
A top quark mass measurement in the \ttbar{} all-jets final state is presented, resulting in
$\mtop = 172.34\pm0.20\statJSF\pm0.76\syst\GeV$.
The modeling of color reconnection provides the largest systematic uncertainty, which is increased with respect to measurements
reported by the CMS Collaboration at $\sqrt{s}=7$ and $8\TeV$, due to the usage of newly available models.
The result is in good agreement with previous measurements.

\end{document}